# Numerical Studies of Optimization and Aberration Correction Methods for the Preliminary Demonstration of the Parametric Ionization Cooling (PIC) Principle in the Twin Helix Muon Cooling Channel

J.A. Maloney[a,c,f], V.S. Morozov[b], Ya. S. Derbenev[b], A. Afanasev[c,d], R.P. Johnson[c], C.A. Ankenbrandt[c], C. Yoshikawa[c], K. Yonehara[e], D. Neuffer[e], B. Erdelyi[f],

[a]Triumf, Vancouver, B.C., V6T2A3, CANADA
[b]Thomas Jefferson National Accelerator Facility, Newport News, Virginia 23606, USA
[c]Muons, Inc., Batavia, Illinois 60510, USA
[d]The George Washington University, Washington, D.C. 20052, USA
[e]Fermi National Accelerator Laboratory, Batavia, Illinois 60510, USA
[f]Northern Illinois University, DeKalb, Illinois 60115, USA

Muon colliders have been proposed for the next generation of particle accelerators that study high-energy physics at the energy and intensity frontiers. In this paper we study a possible implementation of muon ionization cooling, Parametric-resonance Ionization Cooling (PIC), in the twin helix channel. The resonant cooling method of PIC offers the potential to reduce emittance beyond that achievable with ionization cooling with ordinary magnetic focusing. We examine optimization of a variety of parameters, study the nonlinear dynamics in the twin helix channel and consider possible methods of aberration correction.

## I. INTRODUCTION

Muon colliders have been proposed for the next generation of particle accelerators that study high-energy physics at both the energy and intensity frontiers. One of the principal technical challenges in designing a next generation muon collider is muon beam cooling. Muons are produced as tertiary particles, and a muon beam will occupy a very large volume of phase space. Muon cooling reduces the emittance of the beam, a measure of the phase space volume of the beam. This can improve both the luminosity and the physics reach of the collider. A muon collider will need to reduce beam emittance by approximately 6 orders of magnitude to fit within the dynamic aperture of accelerating structures and meet collider luminosity goals [1]. The need for precise energy resolution also makes muon cooling essential for a muon-based Higgs factory operating at the intensity frontier [2]. Additional transverse beam cooling as supplied by the PIC method described here allows other benefits such as even higher luminosity, reduced detector backgrounds and reduced demands on the proton driver and accelerating systems. [3]

To achieve a collider in the luminosity range from $10^{34}$-$10^{35}$ cm$^{-2}$ s$^{-1}$, the transverse emittance for the muon beams must be reduced to about 10 $\pi$ mm-mr. Conventional beam cooling techniques are not useful for cooling such muon beams. The cooling timescales for laser cooling (~$10^{-4}$ s), stochastic cooling (many seconds) and electron cooling (~$10^{-2}$ s), for example, are too long compared to the short muon lifetime of 2.2 μs. The large muon mass also means that synchrotron radiation cooling, an effective technique for electron/positron beams, cannot be used in a muon collider. Instead, muon cooling channels reduce emittance through ionization cooling. One proposal for the final stage of 6D muon cooling is a channel that utilizes the principle of parametric-resonance ionization cooling (PIC) and is based on a twin helix [4]. This channel couples ionization cooling with induced ½ integer resonances to achieve strong focusing in the beam. The potential of PIC is a reduction in equilibrium emittance in each transverse plane by about a factor of 10, which would increase luminosity by a factor of 10 beyond the level of non-resonant ionization cooling using the same magnetic field strengths. Control of nonlinear dynamics is needed to achieve the cooling potential of PIC because uncorrected aberrations in the beam optics can overcome the strong focusing effects of the induced resonances.

This paper studies how a variety of parameters of the twin helix channel may be optimized. This paper also details studies, using COSY INFINITY, of the nonlinear dynamics in the twin helix channel and considers possible methods of aberration correction.

## II. PARAMETRIC-RESONANCE IONIZATION COOLING AND THE TWIN HELIX CHANNEL

The principle of parametric-resonance ionization cooling (PIC) can be implemented in a channel where the muon beam is transported in a periodic magnetic structure. The ordinary phase space trajectory for particles in a stable orbit in this channel is elliptical. The magnetic fields in the channel are perturbed to induce a half-integer resonance that alters the phase space trajectories of particles at periodic fixed points in the channel from parabolic to hyperbolic orbits. This results in strong focusing in position at the expense of growth in angular divergence of the beam [5]. A key aspect of PIC is correlated optics where this strong focusing occurs simultaneously in both the horizontal and vertical planes at a location with small, but non-zero dispersion to allow for 6D cooling through emittance exchange [6]. Wedge absorbers are placed at these points to stabilize the growth in angular divergence and enable emittance exchange. RF cavities are interspersed between the absorbers to maintain the reference momentum of the beam.

One proposed implementation of the PIC principle is the twin helix channel [7]. The basic twin helix channel utilizes a pair of superimposed helical magnetic dipole harmonics. These helical harmonics have equal field strength, and equal but opposite helicity. This creates a channel with alternating dispersion that is practically fringe-field-free. The magnetic fields for helical harmonics can be analytically expressed in cylindrical coordinates as:

$$B_\phi^n(\phi,\rho,z) = B_n[I_{n-1}(nk\rho) - I_{n+1}(nk\rho)]\cos(n[\phi - kz + \phi_0^n]) \quad (1)$$

$$B_\rho^n(\phi,\rho,z) = B_n[I_{n-1}(nk\rho) + I_{n+1}(nk\rho)]\sin(n[\phi - kz + \phi_0^n]) \quad (2)$$

$$B_z^n(\phi,\rho,z) = -2B_n I_n(nk\rho)\cos(n[\phi - kz + \phi_0^n]) \quad (3)$$

$$\text{where} \quad B_n = \left(\frac{2}{nk}\right)^{n-1} \frac{\partial^{n-1} B_\phi^n}{\partial \rho^{n-1}}\bigg|_{\substack{\phi+\phi_0^n=0 \\ \rho=0 \\ z=0}} \quad (4)$$

$B_n$ is the appropriate derivative of the magnetic field strength, and $I_n(x)$ is the modified Bessel function of the first kind. The order of the Bessel functions and the harmonic multipole are determined by the parameter, $n$ ($n = 1$ for dipole, $n = 2$ for quadrupole, etc.).

When two helical harmonics with equal field strength and equal but opposite helicity are superimposed, the total horizontal (x) and longitudinal (z) components of the magnetic field vanish and the resulting magnetic field in the x-z mid-plane simplifies to:

$$\vec{B} = 2B_n[I_{n-1}(nkx) - I_{n+1}(nkx)]\cos(n[kz + \phi_0^n])\hat{y} \quad (5)$$

This channel enables a beam of particles to have a stable reference orbit in the x-z mid-plane [8]. Additional pairs of similarly matched helical harmonics, as well as continuous "straight" magnetic multipoles, may be superimposed while maintaining this planar reference orbit. By using combinations of continuous magnetic fields and helical harmonic pairs, the optics of the channel can be adjusted without complications caused by fringe fields from a series of lumped elements with discrete length.

In addition to the helical dipole harmonic pair, a continuous straight quadrupole magnetic field is superimposed onto the basic twin helix channel to redistribute focusing between the horizontal and vertical planes. This basic channel is designed to satisfy the PIC requirement of correlated optics: the horizontal and vertical betatron tunes are both integer multiples of the dispersion function [9]. The parameters of this basic twin helix channel also can be easily rescaled. The scaling relationships in 6-9 are given for arbitrary values of the momentum of the reference particle ($p$) and the period length of the helical dipole harmonic ($\lambda$) as:

$$B_d = 6.515 \cdot 10^{-3} \ [T \cdot m/MeV/c] \ p/\lambda \tag{6}$$

$$\partial B_y/\partial x = 2.883 \cdot 10^{-3} \ [T \cdot m/MeV/c] \ p/\lambda^2 \tag{7}$$

$$x_{max} = 0.121 \ \lambda \ [m] \tag{8}$$

$$D_x = 0.196 \ \lambda \ [m] \tag{9}$$

where $B_d$ is the field strength of each helical dipole harmonic, $\partial B_y/\partial x$ is the gradient of the continuous quadrupole, $x_{max}$ is the maximum amplitude of the periodic orbit in the horizontal mid-plane and $D_x$ is the maximum dispersion amplitude [10].

Two additional pairs of helical quadrupole harmonics are superimposed onto this basic channel to perturb the magnetic fields to induce the PIC ½-integer resonance condition [11]. One pair induces the resonance in the vertical plane while the other induces the resonance in the horizontal plane. These additional harmonics alter each particle trajectory within the beam so that strong focusing reduces spot size at regular intervals throughout the channel at the expense of increased angular divergence. To avoid beam instability caused by these resonances, wedge absorbers are placed at every other point of strong focusing to stabilize the beam and enable ionization cooling and emittance exchange [12]. An RF cavity is placed after each absorber to restore and maintain the beam's reference momentum.

A linear model of this channel has been previously simulated using COSY INFINITY (COSY) [13]. COSY uses differential algebraic techniques to allow computation of Taylor maps for arbitrary order, allowing the user to disentangle linear and non-linear effects [14]. Additional coding was added to modify the COSY beam physics package and include the stochastic effects of multiple Coulomb scattering and energy straggling in these simulations [15]. Beam cooling has been demonstrated with this non-optimized model [16]. A comparison was made of the cooling effects of the channel with and without inducing the PIC resonance. These simulations verified the analytic theory, showing improved cooling with PIC [17].

### III.  OPTIMIZATION OF THE TWIN HELIX CHANNEL

Optimization of the channel design began with consideration of the period length for the helical dipole harmonics. The final stage of 6D cooling in a muon collider or Higgs factory needs to reduce emittance by about 2 orders of magnitude beyond the reduction achieved through initial 6D cooling methods. The drift space between wedges should be minimized to maximize ionization cooling per channel length. Additionally, the total length of this final stage of 6D cooling should be as short as possible to minimize losses due to muon decays. Consideration must also be given, however, to allow sufficient space for energy-restoring RF cavities.

Using the scaling relationships for the twin helix, eqns. 6-9, the parameters for the basic twin helix channel can be adjusted for different helical dipole period lengths. The maximum amplitude and maximum dispersion can also be determined. For a variety of helical dipole period lengths, G4Beamline (G4BL) [18], was used to track particle distributions and the transmission rate was measured after 1000 absorbers. The simulations tracked a distribution of 1000 muons with momentum of 250 MeV/c. These simulations indicate that choosing a helical dipole period of approximately 20 cm was optimal for improved transmission while minimizing the drift space between absorbers. Table I shows the scaling relations among the helical dipole period, magnetic field strengths, maximum offset, dispersion, and the resulting transmission rate.

TABLE I.    Comparison of parameters for various twin helix configurations.

| $\lambda_D$ | $B_D$ | $\partial B_y / \partial x$ | $x_{max}$ | $D_{x\,max}$ | % transmission |
|---|---|---|---|---|---|
| .05 cm | 48.84 T | 76.08 T/m | 0.006 cm | 0.010 m | 03.0 |
| .10 cm | 24.42 T | 38.04 T/m | 0.012 cm | 0.020 m | 26.1 |
| .20 cm | 16.28 T | 18.02 T/m | 0.024 cm | 0.039 m | 37.3 |
| .30 cm | 08.14 T | 09.01 T/m | 0.036 cm | 0.059 m | 35.9 |

This choice of 20 cm for the period of the helical dipole harmonics has additional advantages. The maximum dispersion is just under 4 cm, which is approximately the same small, but non-zero, value at the location chosen for the wedge absorber in the prior simulations with a helical dipole period of one meter [19]. For a twin helix channel with a 20 cm dipole period, wedge absorbers would be ideally placed at the points of maximum dispersion. As illustrated in Fig. 1, this allows RF cavities to be symmetrically placed between the absorbers.

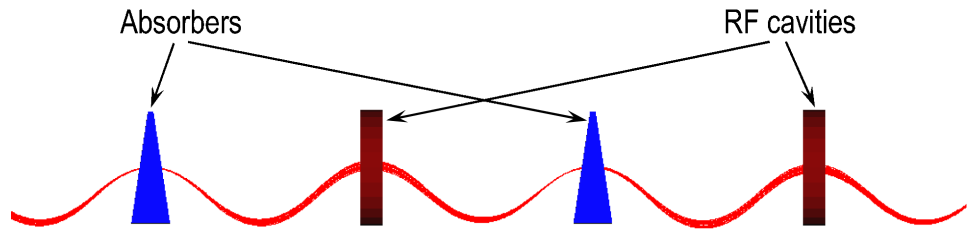

FIG. 1. (Color) Schematic of symmetric twin helix layout.

In addition to improving transmission, shortening the helical dipole period also reduces some beam aberrations. One major drawback, however, is the increase in the strengths of the magnetic fields as the helical dipole period is reduced. For a 20 cm helical dipole period, the combined field from the pair of helical dipole harmonics scales to 16.3 T. With a 20 cm period for the helical dipole harmonic in the basic twin helix channel, wedge absorbers placed every other period allow less than 40 cm for the RF cavities. The result is a set of parameters that are challenging but within the levels contemplated for an energy frontier muon collider. The 20 cm period maximizes transmission while minimizing unnecessary length between the wedge absorbers, providing a good balance between cooling efficiency and channel length. The dipole period can be rescaled, if necessary, to an increased length to reduce these magnetic field strengths.

After selecting the period length for the helical dipole harmonics, the simulations of the linear model of the channel were repeated using COSY. For a 20 cm period, the field strength of each helical dipole harmonic was 8.14375 T, with a superimposed continuous quadrupole field of 18.01875 T/m. A test particle was tracked every 40 cm (2 helical dipole periods) as it traveled in this channel. Relative to the reference orbit, the particle follows an elliptical trajectory as shown in Fig. 2.

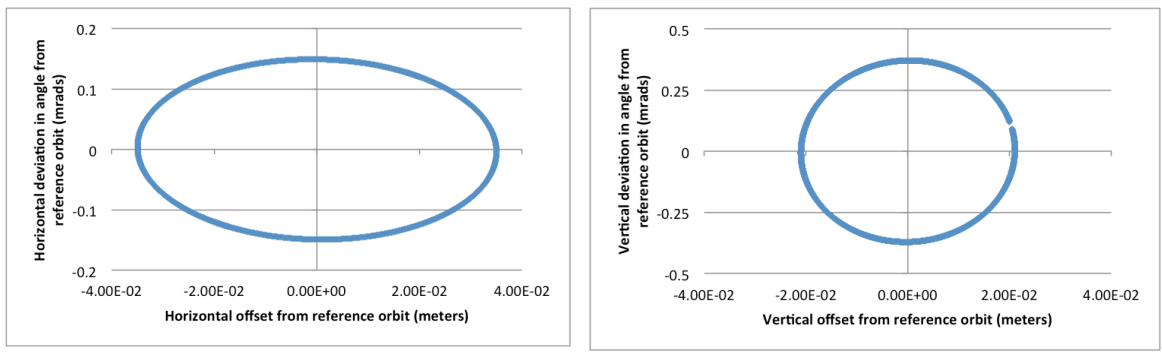

FIG. 2(a)-(b). (Color) The $\lambda_D$=20 cm basic twin helix channel simulated in COSY without wedge absorbers or energy restoring RF cavities. The trajectory of a 250 MeV/c $\mu^-$ was launched offset in both planes from the reference orbit by 2 cm and 130 mr and tracked and plotted every two dipole periods in (a) the horizontal plane and (b) vertical planes.

To induce the horizontal ½ integer resonance, a pair of helical quadrupole harmonic fields was added to the basic channel. Each harmonic in this pair had a field strength of 0.063662 T/m, a period of 80 cm and a phase advance that shifted the location of the maximum field amplitude by 30 cm relative to maximum field for the underlying helical dipole harmonic pair. The vertical resonance was induced by an additional pair of helical quadrupole harmonics, which each had a field strength of 0.127324 T/m, a period of 80 cm and a 4.4 cm phase shift. Fig. 3 shows the change in trajectory for the test particle once the PIC resonances have been introduced.

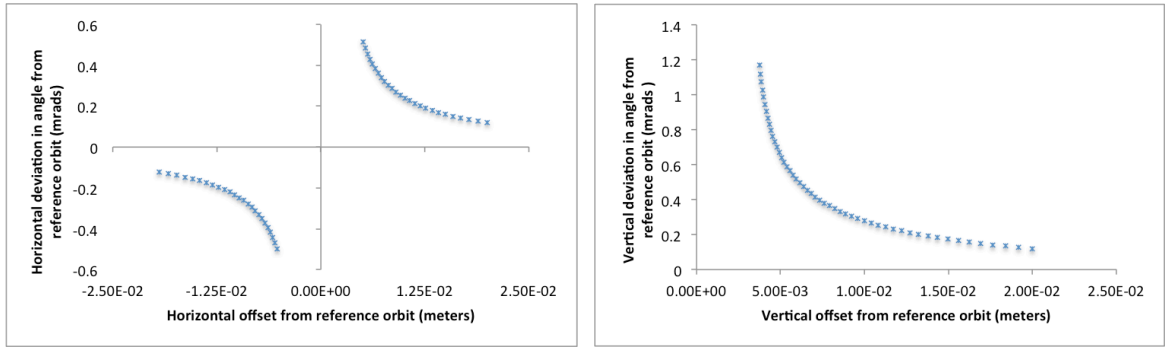

FIG. 3(a)-(b). (Color) The $\lambda_D$=20 cm basic twin helix channel with induced ½-integer parametric resonance simulated in COSY without wedge absorbers or energy restoring RF cavities. The trajectory of a 250 MeV/c $\mu^-$ was launched offset in both planes from the reference orbit by 2 cm and 130 mr and tracked and plotted every two dipole periods in (a) the horizontal plane and (b) vertical planes.

The resonance can also be seen in the linear transfer matrix for one cell in this channel. Under this approach, the horizontal and vertical magnification matrix elements (x|x) and (y|y) should be less than one, while (x|a) and (y|b) should approach zero, where a = $p_x/p_0$ and b = $p_y/p_0$. In the present model, for example, the values for these matrix elements are presented in Table II.

TABLE II. The 4x4 sub-matrix for the $\lambda_D$=20 cm basic twin helix cell with induced parametric resonance.

|     | (x) | (a) | (y) | (b) |
| --- | --- | --- | --- | --- |
| (x) | -0.971 | -6.26 x $10^{-3}$ | 0 | 0 |
| (a) | 5.70 x $10^{-5}$ | -1.03 | 0 | 0 |
| (y) | 0 | 0 | 0.958 | 1.01 x $10^{-4}$ |
| (b) | 0 | 0 | 8.26 x $10^{-2}$ | 1.04 |

Wedge absorbers were added at every other periodic focal point, with RF cavities symmetrically placed in between them. The central thickness of the wedge absorbers was arbitrarily set at 2 cm. The frequency and phase of the RF cavities were chosen as 201.5 MHz and 30 degrees. This is similar to cavities used in simulations for a helical cooling channel (HCC) [20] that could be upstream of a twin helix final cooling channel.

Optimization of the wedge thickness and RF parameters are not being studied at this time. In the final channel design, wedge thickness will be decreased as the beam travels through the channel to maintain cooling efficiency [21]. Parameters for the RF cavities will also need to be adjusted to maintain the reference momentum and match the time structure of the beam. Optimization of these parameters is left for consideration until after the exact structure of the upstream beam being delivered to the PIC channel and the beam acceptance parameters for the downstream accelerating structures have been determined.

The effects of varying the horizontal wedge gradient (dx/dz) were studied without a resonance (no helical quadrupole harmonic pairs). Fig. 4 shows some examples with (a) a flat absorber, (b) 0.10 wedge gradient, (c) 0.20 wedge gradient, and (d) 0.30 wedge gradient with the wedge orientation reversed. Increasing the gradient increases the rate of longitudinal cooling but also reduces horizontal transverse cooling by the same rate [22]. The final example demonstrates the principle of reverse emittance exchange (REMEX) used to gain extra reduction in transverse emittance at the expense of an increase in longitudinal emittance [23]. A wedge gradient of 0.10 was chosen to balance transverse cooling with longitudinal cooling through emittance exchange.

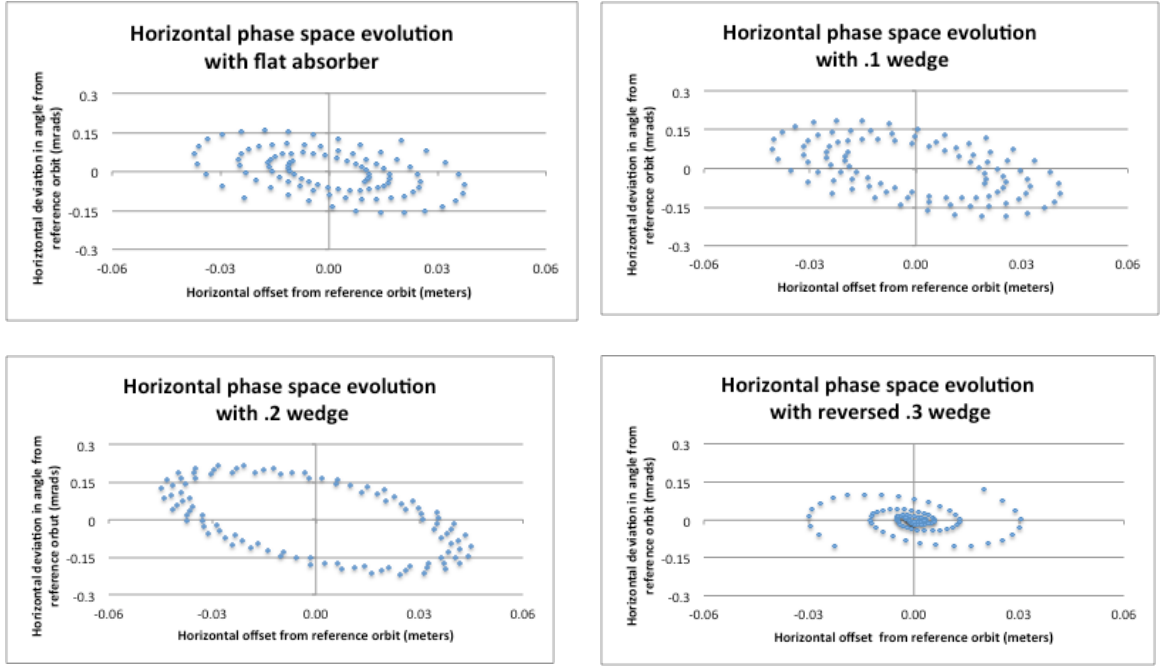

FIG. 4(a)-(d). (Color) The $\lambda_D$=20 cm twin helix channel simulated in COSY with wedge absorbers and energy restoring RF cavities. The trajectory of a 250 MeV/c $\mu^-$ was launched offset in both planes from the reference orbit by 2 cm and 130 mr and tracked and plotted every two dipole periods in the horizontal plane for (a) flat absorber; (b) 0.10 wedge gradient; (c) 0.20 wedge gradient; and (d) 0.30 wedge gradient with reverse wedge orientation (REMEX).

Next, the helical quadrupole harmonic pairs were added to induce the PIC resonance for the chosen wedge gradient. The effects of varying the field strength of the resonance harmonic pairs were studied. The same field strength of 0.127324 T/m used to induce the resonance in the vertical plane, shown in Fig. 3(b), was maintained. Fig. 5 shows the trajectory of the test muon in the vertical plane with this resonance induced.

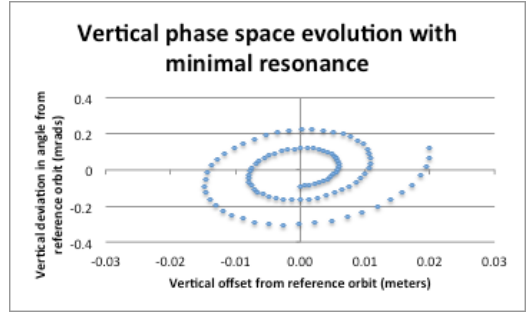

FIG. 5. (Color) The $\lambda_D$=20 cm full twin helix channel with induced vertical resonance simulated in COSY. The trajectory of a 250 MeV/c $\mu^-$ was launched offset in both planes from the reference orbit by 2 cm and 130 mr and tracked and plotted every two dipole periods in the vertical plane.

The field strength of the helical quadrupole harmonic pair used to induce horizontal phase space resonance in the basic channel, as shown in Fig. 3(a), was not strong enough once the wedge and RF cavities were added. As a result, the magnification term in the linear transfer matrix, (x|x), was greater than 1. This effect is caused by a parasitic resonance that has been previously identified [24]. To induce the resonance, the strength of this helical harmonic pair was doubled. Cooling simulations showed that increasing the field strength further distorted the beam and reduced the cooling efficiency of the channel. Fig. 6 shows the effects on the trajectory of a test particle in horizontal phase space with (a) no resonance, (b) below resonance (0.063662 T/m), (c) resonance minimally triggered (0.127324 T/m), and (d) resonance strongly triggered (0.31831 T/m).

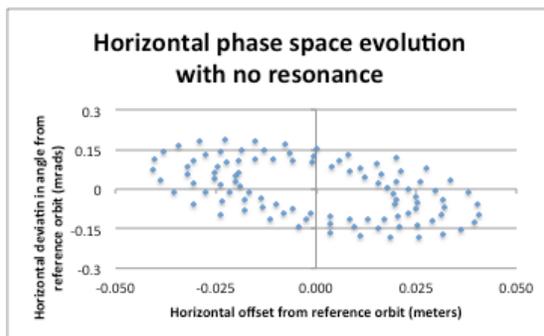
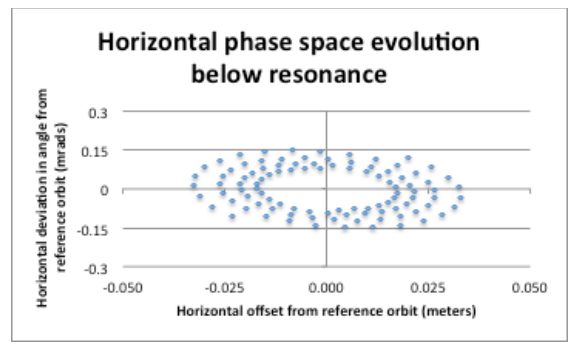

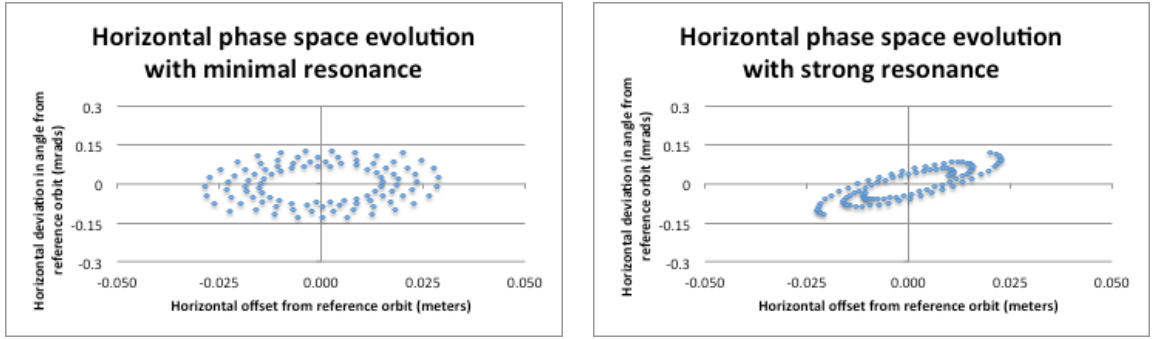

FIG. 6(a)-(d). (Color) The $\lambda_D$=20 cm full twin helix channel with induced horizontal resonance simulated in COSY. The trajectory of a 250 MeV/c $\mu^-$ was launched offset in both planes from the reference orbit by 2 cm and 130 mr and tracked and plotted every two dipole periods in the horizontal plane with (a) no resonance induced, and (b)-(d) various strengths for resonance-inducing harmonic fields.

After linear simulations of cooling without stochastic effects for a distribution of particles, it was determined that using the parameters listed in Table III provided the best cooling efficiency after 200 cells. The initial distribution had a 6D emittance of $1.27 \times 10^{-9}$ m$^3$-rad$^2$. Without inducing the parametric resonance, after 200 cells, 6D emittance was reduced to $7.93 \times 10^{-14}$ m$^3$-rad$^2$. With the induced resonance, 6D emittance was reduced to $4.39 \times 10^{-14}$ m$^3$-rad$^2$ after the same number of cells. Additional optimization of the wedge thickness and RF cavity parameters is expected to improve this result and improve cooling by more than merely a factor of 2.

TABLE III. Parameters for optimized twin helix cell.

| PARAMETER | VALUE |
|---|---|
| Reference particle | 250 MeV/c $\mu^-$ |
| H. Dipole field | 8.14375 T |
| H. Dipole wavelength | 20 cm |
| Straight Quadrupole field | 18.01875 T/m |
| H. Quadrupole field (horizontal resonance pair) | .4/p T/m |
| H. Quadrupole wavelength | 80 cm |
| H. Quadrupole phase advance | 30 cm |
| H Quadrupole field (vertical resonance pair) | .4/p T/m |
| H. Quadrupole wavelength | 40 cm |
| H. Quadrupole phase advance | 4.4 cm |
| Beryllium wedge central thickness | 2 cm |
| Wedge angle gradient | .10 |
| RF cavity voltage | -12.546 MV |
| RF frequency | 201.25 MHz |
| RF phase | 30 degrees |

Next, stochastic effects were added to the simulations. This was done through modification of the COSY language to calculate and apply a unique stochastic map for each particle interacting with any absorber, following the same methodology used to verify PIC theory for the linear model [25]. Fig. 7(a) and (b) shows the results of the tracking of the trajectory for an individual test muon in the horizontal and vertical phase space for 200 cells with the addition of stochastic effects of multiple Coulomb scattering and energy straggling.

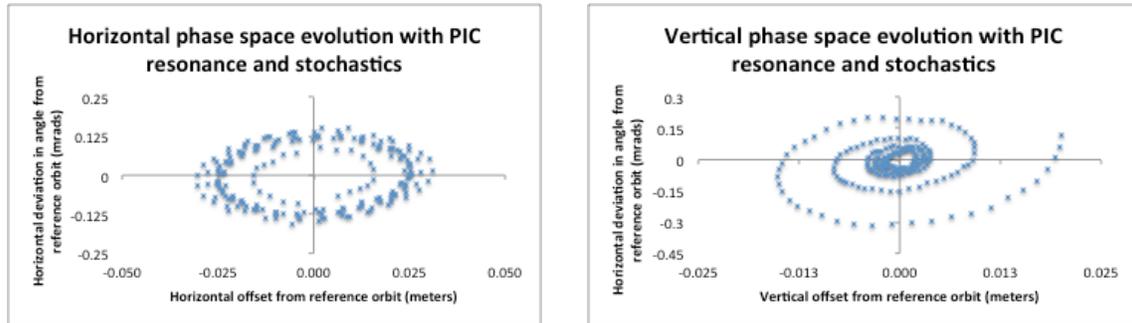

FIG. 7(a)-(b). (Color) The optimized twin helix channel simulated in COSY with the stochastic effects of multiple Coulomb scattering and energy straggling. The trajectory of a 250 MeV/c μ- was launched offset in both planes from reference orbit by 2 cm and 130 mr and tracked at the center of each wedge absorber in the (a) the horizontal, and (b) vertical planes.

The parameters given in Table III were used for a distribution of 1000 test muons with stochastic effects of multiple Coulomb scattering and energy straggling. For this distribution, the initial coordinates were calculated using a Gaussian distribution with the following sigma for deviations from the reference orbit: (1) offset in each plane: 2 mm; (2) offset in angle in each plane: 130 mr; (3) energy spread: 1%; and (4) longitudinal bunch length: 2 cm. These beam parameters were chosen based on the expected output from an upstream initial 6D helical cooling channel. The results are shown in Fig. 8-10. Without optimizing wedge thickness or the RF parameters, equilibrium is reached after about 400 cells. The channel's target of a reduction of 6D emittance by 2 orders of magnitude is accomplished after only about 100 cells, equating to a channel length of approximately 40 meters.

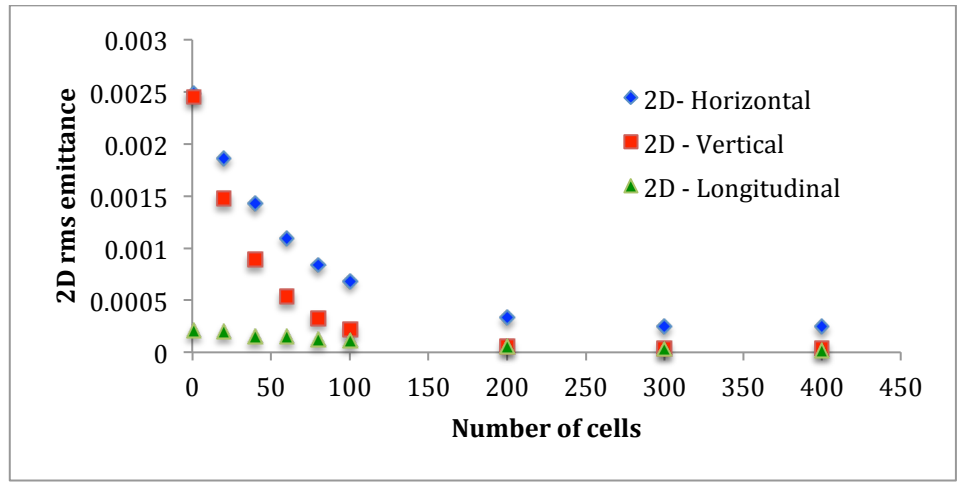

FIG. 8. (Color) Reduction in 2D emittance as a function of channel length.

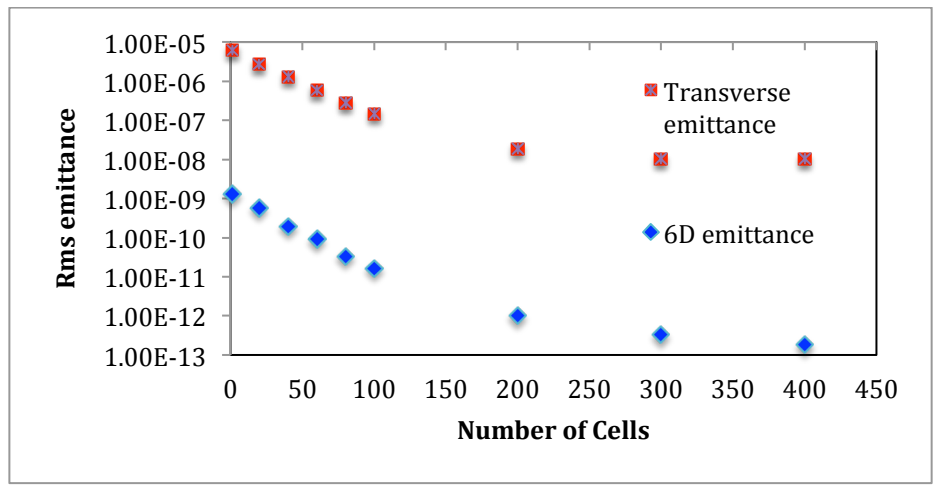

FIG. 9. (Color) Changes in transverse and 6D rms emittance as a function of channel length.

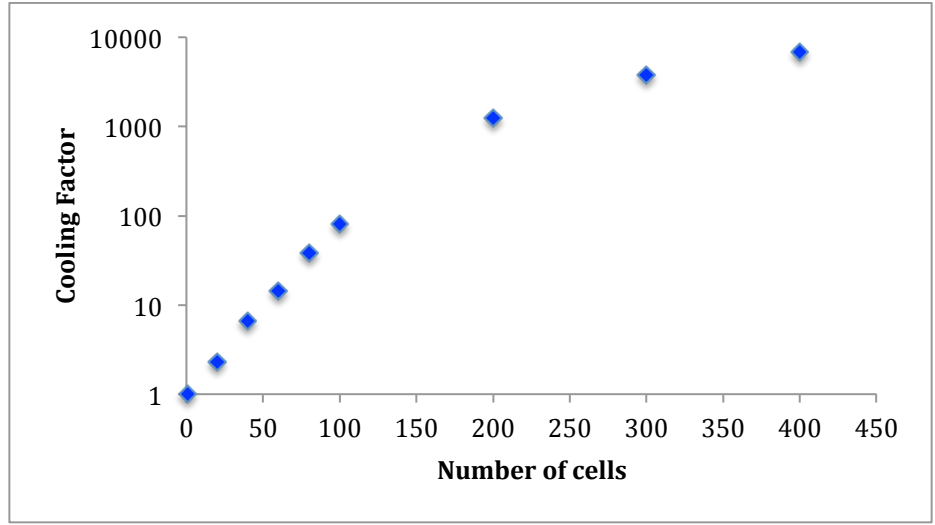

FIG. 10. (Color) Cooling factor (initial 6D emittance/final 6D emittance) as a function of channel length.

## IV. EVALUATION OF ABERRATIONS IN THE TWIN HELIX CHANNEL

Having studied methods to optimize various parameters for the twin helix channel and simulated cooling with a linear model, the effects of aberrations in the channel need to be evaluated. The linear model described above provides an important tool for optimizing the PIC cooling channel design. This linear model simulates the efficiency of the cooling channel if all non-linear aberrations in the system are perfectly corrected [26]. Since muon beams can have a very large initial angular and energy spread, non-linear effects (aberrations) in the system dependent on these parameters can dramatically change the final spot size of the beam. The efficiency of aberration correction can be determined by comparing the corrected system, including non-linear effects, against the linear model.

The optimized PIC channel was simulated to determine how aberrations impacted performance of the cooling channel. The transfer and aberration maps are calculated by COSY from a point on the reference particle orbit at the center of one wedge absorber to the point on the reference orbit that is the center of the next wedge absorber based on the maximum beam size parameters. The optimized twin helix parameters from Table III and a reference momentum of

250 MeV/c were used. Aberrations were determined for: (1) a horizontal and/or vertical deviation from the reference orbit of up to 2 mm in position; (2) a horizontal and/or vertical deviation from the reference orbit in angle of up to 130 mr; (3) a bunch length of up to 3 cm; and (4) a deviation in momentum ($\Delta p/p$) of up to 2%.

Table IV lists the largest $2^{nd}$ and $3^{rd}$ order aberrations affecting the final spot size at the periodic focal points in the channel. The maximum effect on final spot size for all other $2^{nd}$ and $3^{rd}$ order aberrations is less than 1 mm. The aberration (x|aa), for example, shows the variation (in meters) in final horizontal position of the particle as a function of the square of its initial angle ($p_x/p_0$) in the horizontal plane.

TABLE IV. Largest $2^{nd}$ and $3^{rd}$ order aberrations affecting final spot size for the optimized basic and full twin helix channel with $\lambda_D$=20 cm and a 250 MeV/c reference muon.

| Aberration | Full Cell Parameter (meters) | Basic Cell Parameter (meters) |
|---|---|---|
| (x\|aa) | 0.00235 | 0.00173 |
| (x\|a$\delta$) | 0.00218 | 0.00208 |
| (x\|aaa) | -0.01760 | -0.01920 |
| (x\|abb) | -0.00599 | -0.00640 |
| (y\|aab) | 0.00598 | 0.00650 |
| (y\|bbb) | 0.00111 | 0.00122 |

The data in Table IV is noteworthy in that the aberrations for the basic channel are identical in nature and nearly identical in size to those for the full twin helix channel, with resonance inducing helical harmonic pairs, wedge absorbers and RF cavities. Since the aberrations are due primarily to the optics of the basic channel's helical dipole pair and straight quadrupole components, and not the additional elements in the full channel, aberration correction efforts can focus initially on correcting aberrations without the complications of the induced resonances, wedges and RF components.

The aberration maps calculated with COSY show that angular-based aberrations have the greatest effect on the final position of particles in the channel. Because the angular spread in the muon beam can be large, correcting these aberrations is critical to a successful cooling channel design. In addition to the 2nd and 3rd order aberrations previously noted, angular aberrations at 5th and 7th order were also non-trivial. In particular, (x|aaaaa) and (x|aaaaaaa), the aberrations to the final horizontal offset from reference orbit based on 5th and 7th powers of the initial horizontal angular deviation from the reference orbit ($p_x/p_0$), caused substantial instability and particle loss.

Correction efforts focused on superimposing a variety of continuous magnetic fields on the channel. A series of simulations were performed to study the effects of various parameters on the largest 2nd and 3rd order aberrations as well as other aberrations increased by adding correcting magnetic fields.

The basic twin helix channel was simulated with continuous sextupole, octupole and decapole fields superimposed onto it one at a time. The effects on the identified aberrations were studied as the pole tip field strengths of these multipoles were varied. Fig. 11 shows an example of this evaluation for the continuous octupole field. The effects on aberrations are plotted in Fig. 11 as functions of the field strength parameter.

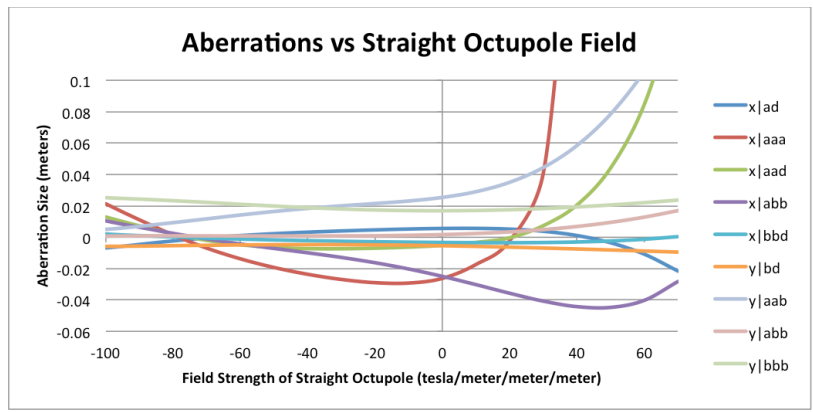

FIG. 11. (Color) Dependence of twin helix aberrations on the continuous octupole magnetic field.

The effect of superimposing various pairs of helical harmonic magnetic fields was also assessed. These included pairs of helical quadrupole, sextupole, octupole, and decapole harmonics. The effects of each pair were studied independently. Like the helical dipole harmonic pair in the basic twin helix channel, each harmonic in these higher order pairs had equal field strengths, phase offsets and equal but opposite helicities. For each superimposed harmonic pair, the effects of varying field strength, wave number, and phase offset on the target aberrations were independently assessed. Figs. 12-14 show examples of this study plotted for the helical sextupole harmonic pairs. Variations in field strength are plotted in a manner similar to that used for the continuous correcting fields. Wavelength variation is plotted as a function of "nk," where n is the number of periods in the given wavelength and k is the wave number ($2\pi/\lambda$).

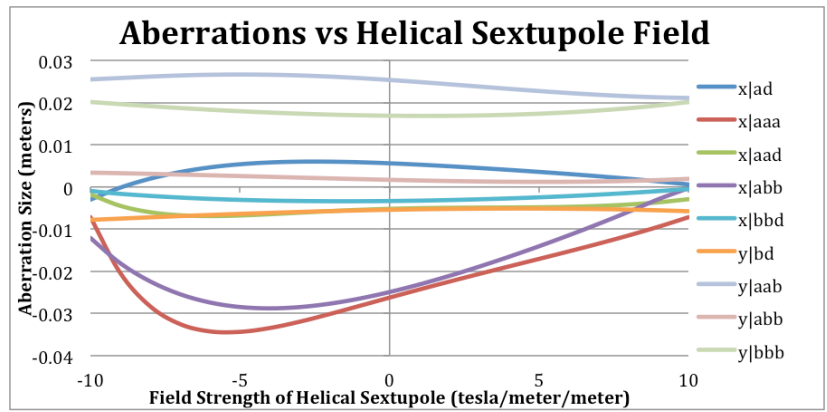

FIG. 12. (Color) Dependence of twin helix aberrations on the helical sextupole harmonic magnetic field. Phase offset from helical dipole pair = 0 and nk = 1.

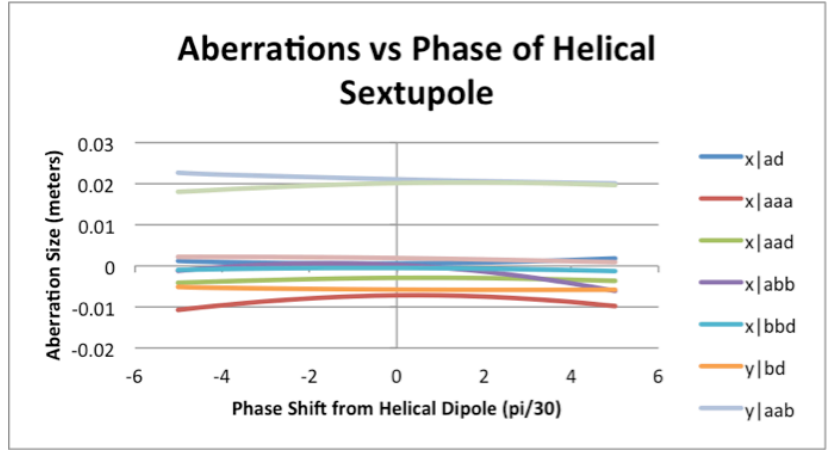

FIG. 13. (Color) Dependence of twin helix aberrations on the helical sextupole harmonic phase offset. Field strength = 10 T/m$^2$ and nk = 1.

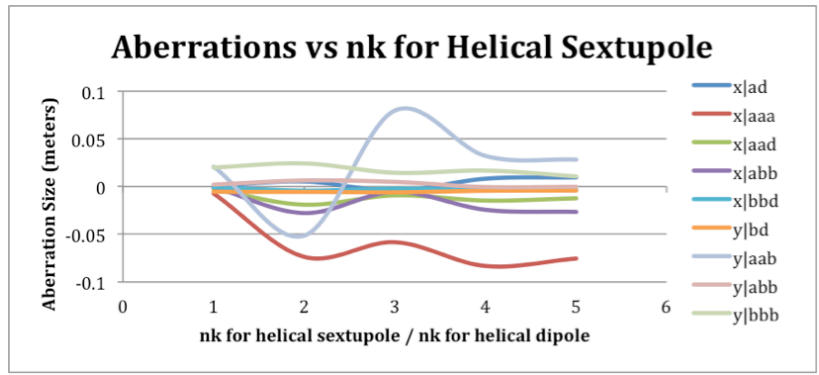

FIG. 14. (Color) Dependence of twin helix aberrations on the helical sextupole harmonic wave number. Field strength = 10 T/m$^2$ and phase offset from helical dipole pair = 0.

These studies identified potential methods for correcting higher order aberrations, as well as which aberrations were least sensitive to correction. Based on what was learned, various combinations of correcting magnetic fields were used in an attempt to minimize aberrations in the basic twin helix channel.

In all cases, the correlated optics condition was maintained, and this means that the reference orbit had to be recalculated since these higher order magnetic fields have a "feed-down" effect that modifies the original orbit of the reference particle. For example, as the reference particle oscillates around the channel's optical axis in the x-z mid-plane, it experiences a dipole-like and quadrupole-like field from a sextupole field that has been superimposed to correct aberrations. Field strength, phase offset, period and harmonic number provide a number of variable parameters for correcting the system.

A correction model was developed after attempts to simultaneously minimize all large aberrations. This design superimposed two additional pairs of helical quadrupole harmonics, one pair of helical sextupole harmonics, and three pairs of helical octupole harmonics onto the basic twin helix channel. Figs. 15 and 16 show that even after one cell, the sensitivity to large initial angle can be seen. In these simulations, four concentric cones of muons are launched from the same position along the reference orbit with the same momentum, 250 MeV/c. These cones deviate from the reference orbit by an angle of $\pm 30, 60, 90$ and $120$ mr. The simulations included all non-linear effects up to $9^{th}$ order, which was sufficient for numerical convergence.

After one cell, as shown in Fig. 15, there is strong focusing for some of the particles, but substantial deviation in final position of others. The muons showing deviation are the ones from the two cones with the largest initial angular deviations from the reference orbit. After 20 cells, as shown in Fig. 16, the surviving muons have the strong focusing characteristic of the PIC resonance effect. Unfortunately, only muons in the cones with angular deviations from the reference orbit of $\pm 30$ and $60$ mr have survived.

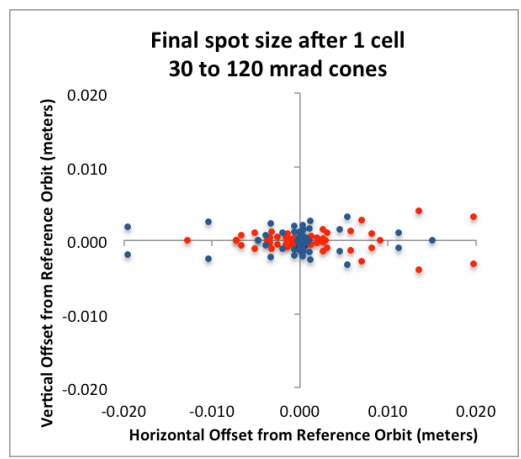

FIG. 15. (Color)  Tracking of concentric cones, with angular deviation of 30, 60, 90 and 120 mr, of 250 MeV/c muons launched on the reference orbit in COSY with non-linear effects through 9th order for the uncorrected (red) and corrected (blue) twin helix channel.

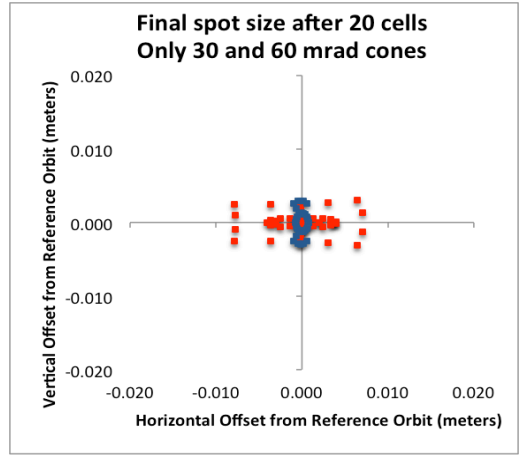

FIG. 16. (Color)  Tracking of two concentric cones, with angular deviation of 60 and 120 mr, of 250 MeV/c muons launched on the reference orbit in COSY with non-linear effects through 9th order for the uncorrected (red) and corrected (blue) twin helix channel.

Fig. 17 displays results of a G4BL simulation of the same channel, tracking muons with initial angular spread of up to ± 70 mr through two helix periods.  Although these results are promising for this small angle distribution, it represents only about half of the equilibrium rms angle expected for a muon with momentum of 250 MeV/c.  Muons with an initial angular deviation greater than about 70 mr are eventually lost.

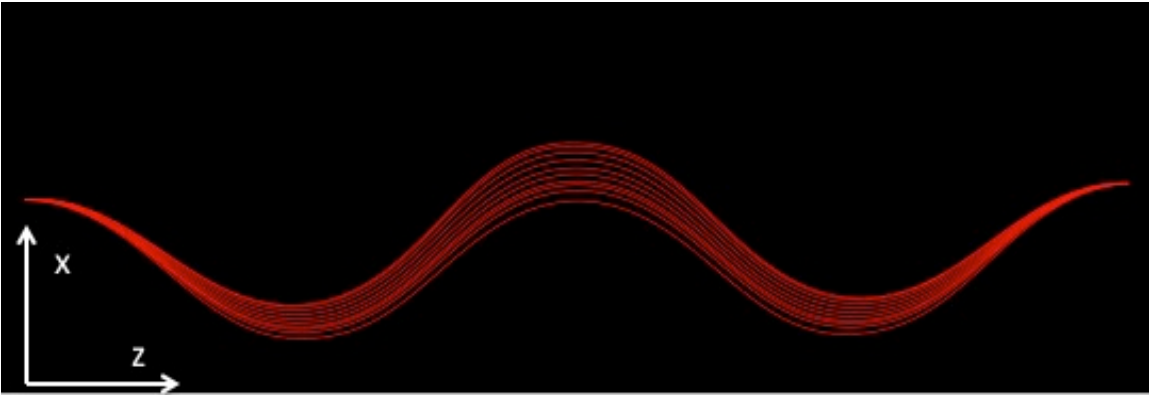

FIG. 17. (Color) G4Beamline simulation of horizontal motion in the x-z mid-plane through 1 cell (2 dipole periods) for a distribution of 250 MeV/c muons launched from the reference orbit with horizontal angular deviations from the reference orbit of up to ±70 mr.

Under the analytic linear model for PIC, the equilibrium beam conditions can be determined as a function of key parameters. The equilibrium rms beam size $\sigma_a$ and angular spread $\theta_a$ at the absorber and the equilibrium rms momentum spread $\Delta p/p$ are given by eqns. 10 [27],

$$\sigma_a^2 = \frac{1}{8} \frac{(Z+1)}{\gamma \beta^2} \frac{m_e}{m_\mu} w^2, \quad \theta_a^2 = \frac{3}{2} \frac{(Z+1)}{\gamma \beta^2} \frac{m_e}{m_\mu}, \quad \left(\Delta p / p\right)^2 = \frac{3}{8} \frac{(\gamma^2+1)}{\gamma \beta^2} \frac{m_e}{m_\mu} \frac{1}{\log} \quad (10)$$

where $\gamma$ and $\beta$ are the usual relativistic factors, $w$ is the wedge absorber central thickness, $m_e$ and $m_\mu$ are the electron and muon masses, respectively, and *log* is the Coulomb logarithm of ionization energy loss for fast particles. Based on these relations, the equilibrium angular and momentum spreads are determined by the reference energy of the muon beam, and the thickness of the wedge absorbers, $w$, can be used to scale the equilibrium beam size.

The dynamic aperture of the corrected system was still smaller than the equilibrium angular spread for a 250 MeV/c muon beam. Since equilibrium angle spread is inversely proportional to the square root of energy, increasing the reference momentum to 1 GeV/c, lowers the equilibrium angle spread from 130 mr to 65 mr. This is within the acceptance shown in Fig. 17. This would mean accelerating the beam after initial 6D cooling and before a final 6D PIC cooling channel. To verify this, the magnet strength and other parameters were scaled and refit for a beam with a 1 GeV/c reference momentum. Fig. 18 shows the results of this higher momentum model with aberration correction. In this model, concentric cones of 1 GeV/c muons deviating in angle from the reference orbit by ±20, 40, 60 and 80 mr, are launched from reference orbit and their final positions are tracked after 40 helical dipole periods. Due to the increase in the reference momentum, this model's dynamic aperture exceeds the equilibrium values for angular spread in the beam. Fig. 19 shows transmission for the same set of concentric cones of 1 GeV/c muons to a final 2 mm spot size after 40 dipole periods. In the uncorrected system, the transmission of muons to a final 2 mm spot size was less than 23%, but in the corrected system this is increased to 60%.

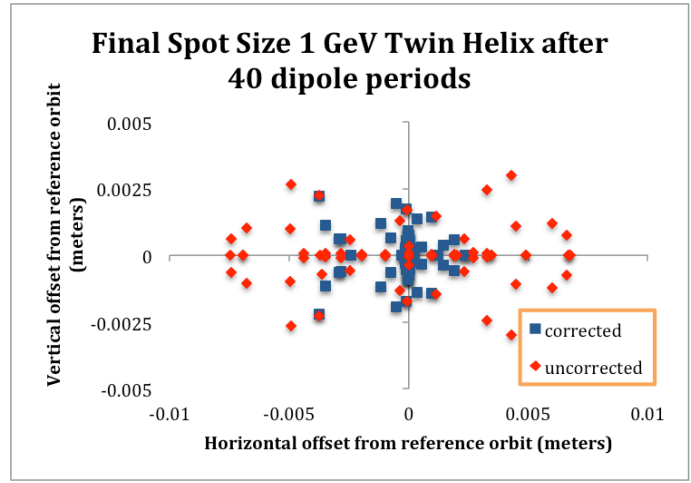

FIG. 18. (Color) Tracking of concentric cones, with initial angular deviation of 20, 40, 60 and 80 mr, of 1 GeV/c muons launched on the reference orbit in COSY with non-linear effects through 5th order for the uncorrected (red) and corrected (blue) twin helix channel.

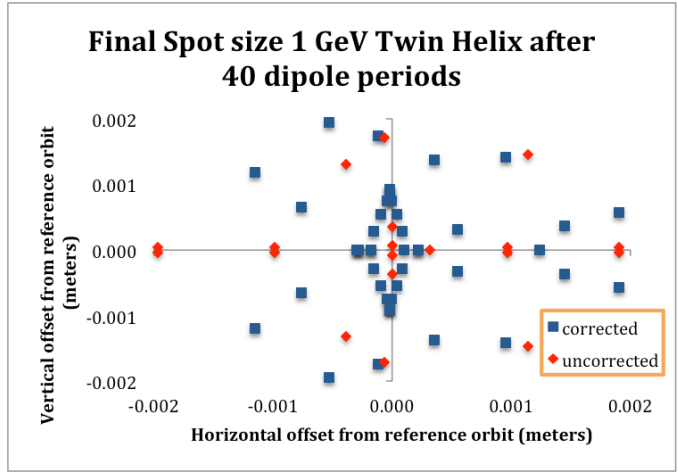

FIG. 19. (Color) Surviving particles within a spot size deviating up to 1 mm from the reference orbit after 40 helical dipole periods for concentric cones, with initial angular deviations of 20, 40, 60 and 80 mr, of 1 GeV/c muons launched on the reference orbit in COSY with non-linear effects through 5th order for the uncorrected (red) and corrected (blue) twin helix channel.

Additional simulations added wedge absorbers to attempt to evaluate cooling in the corrected 1 GeV/c model. Unfortunately, the higher beam energy increased stochastic heating in the beam since energy straggling scales with $\gamma^2$, and this effect overcame the ionization cooling provided by the thin wedge absorbers. Alternative beamline configurations that implement the PIC principle are being investigated to find a cooling channel lattice with less susceptibility to nonlinear effects.

## V. CONCLUSIONS

Muon cooling poses one of the key technical challenges in the successful development of a muon collider to explore high-energy physics at the energy frontier. The same technology will be essential to the development of a muon-based Higgs factory to study physics at the intensity frontier. A twin helix channel implementing the principle of PIC has the potential to improve emittance reduction beyond simple ionization cooling methods and is a strong candidate for final 6D cooling in such colliders. Other candidate final cooling techniques include using lithium lenses [28] or extremely high field solenoids that use new high-temperature superconductors operating at low temperatures [29].

The length of the period for the helical dipole harmonics in the basic twin helix channel was optimized; 20 cm was chosen based on reducing the spacing between wedge absorbers to minimize particle loss through decay. Simulations in COSY verified the induced resonances in both horizontal and vertical dimensions. Wedge absorbers and RF cavities can be placed symmetrically in this channel. Field strengths for the helical quadrupole harmonic pairs were adjusted as necessary to trigger the resonance effects with the addition of absorbers and RF. Cooling in this optimized channel was verified with and without stochastic effects.

The effects of chromatic and spherical aberrations in this optimized channel have been studied. This included comparisons of aberrations in both the basic and full twin helix channel. This comparison demonstrated that the aberrations are arising from the basic channel fields, and that correction effort could focus on this simple system that does not have induced parametric resonances, wedge absorbers or RF cavities. Using COSY to determine a map of beam aberrations, the largest aberrations were identified. Simulations compared how various continuous magnetic multipoles and pairs of helical magnetic harmonics affected these aberrations. Using this information, a modified design successfully corrected major aberrations through $9^{th}$ order. This permitted demonstration of the PIC cooling channel with the inclusion of the correction for aberrations. Unfortunately, the angular acceptance of this channel design was only about half of the equilibrium angular spread in the beam for a reference momentum of 250 MeV/c. By increasing the reference momentum to 1 GeV, the equilibrium angular spread in the beam was reduced to a point that was within the angular acceptance of the twin helix channel. Transmission in the twin helix with aberration correction was substantially improved.

## ACKNOWLEDGEMENTS

Funding for this research was supported, in part, through DOE HEP STTR Grant No. DE-SC00005589, "Epicyclic Helical Channels for Parametric-resonance Ionization Cooling."